\documentclass[prb,twocolumn,superscriptaddress,showpacs]{revtex4}
\usepackage{amsmath, amsfonts, amssymb}
\usepackage{bm}
\usepackage{hyperref}
\usepackage{graphicx}
\usepackage{subfigure}

\begin{document}

\def\Re {\mbox{Re}}
\def\Im {\mbox{Im}}
\newcommand{\avg}[1]{\langle#1\rangle}
\newcommand{\odiff}[2]{\frac{\di #1}{\di #2}}
\newcommand{\pdiff}[2]{\frac{\partial #1}{\partial #2}}
\newcommand{\di}{\mathrm{d}}
\newcommand{\ii}{i}
\newcommand{\norm}[1]{\left\| #1 \right\|}
\renewcommand{\vec}[1]{\mathbf{#1}}
\newcommand{\ket}[1]{|#1\rangle}
\newcommand{\bra}[1]{\langle#1|}
\newcommand{\pd}[2]{\langle#1|#2\rangle}
\newcommand{\tpd}[3]{\langle#1|#2|#3\rangle}
\renewcommand{\vr}{{\vec{r}}}
\newcommand{\vk}{{\mathbf{k}}}
\renewcommand{\ol}[1]{\overline{#1}}
\newcommand{\comments}[1]{}

\title{Josephson Current through a Semiconductor Nanowire: effect of strong spin-orbit coupling and Zeeman splitting}
\author{Meng Cheng}
\affiliation{Station Q, Microsoft Research, Santa Barbara, CA 93106-6105}

\affiliation{Condensed Matter Theory Center, Department of Physics,
University of Maryland, College Park, Maryland 20742, USA }

\affiliation{
Kavli Institute for Theoretical Physics, University of California, Santa Barbara, CA, 93106}

\author{Roman M.~Lutchyn}
\affiliation{Station Q, Microsoft Research, Santa Barbara, CA 93106-6105}

\date{\today}

\begin{abstract}

We study coherent transport through a semiconductor nanowire in the presence of spin-orbit coupling and Zeeman splitting due to an applied magnetic field. By employing analytical and numerical techniques we develop a theory for the Josephson effect in the superconductor-semiconductor nanowire-superconductor structure. We show that Josephson current through the clean semiconductor nanowire exhibits a number of interesting features due to the interplay between the Zeeman splitting and spin-orbit coupling. We also study effect how disorder in the nanowire affects Andreev bound-state energy spectrum and calculate local density of states at the junction.

\end{abstract}
\maketitle

\section{Introduction}\label{sec:intro}
Josephson effect is an example of macroscopic quantum phenomena and is of fundamental importance in condensed matter physics. The applications of the Josephson effect are both numerous and profound~\cite{Golubov_RMP'04, Buzdin'05, Bergeret'05}.
Recently, there has been a growing interest in understanding coherent transport phenomena in hybrid heterostructures involving s-wave superconductors and non-superconducting nanostructures~\cite{Franceschi'11, vanDam2006, Katsaros, Dirks'11, Kasumov, Morpurgo, Doh'05, Xiang}. The possibility to control various parameters of the nanostructures allows one to explore different aspects of the Josephson effect in these devices.

In this paper, we study Josephson effect in superconductor-semiconductor nanowire-superconductor heterostructure (SC-SM-SC) shown in Fig~\ref{fig:junction}a. The system consists of a semiconductor nanowire with strong Rashba spin-orbit coupling(SOC) subject to a large in-plane magnetic field. The wire is connected to the superconducting leads as shown in Fig.~\ref{fig:junction}a. When chemical potential is in the gap, see Fig.~\ref{fig:junction}b, there is only one Fermi surface
~\cite{Streda'03}.
Coherent transport through the nanowire in the helical phase, which originates from the non-trivial interplay between large Zeeman splitting and strong Rashba SOC, have not been previously explored~\cite{helic}.
Previous studies of the Josephson effect~\cite{Dimitrova2006, BobkovaPRB2002, Buzdin_JETPL1982, pi-junction} considered the case of either small Zeeman field or small SOC whereas we are interested in a non-perturbative limit where {\it both} Zeeman and SO terms are large leading to a significant change of the nanowire bandstructure. Such heterostructures involving semiconductor nanowires and s-wave superconductors exhibit a number of interesting effects and, in particular, are very promising for realizing topological superconducting phase hosting Majorana zero-energy modes~\cite{Lutchyn'10, Oreg'10}. We note that the present setup is different from the one in Ref.~[\onlinecite{Lutchyn'10}] leading to $4\pi$ ac Josephson effect because the SC parts are topologically trivial and there are no Majorana bound states.

In this paper, we develop a quantitative theory for the coherent transport through the nanowire in the clean limit by employing a combination of analytical and numerical techniques. The supercurrent through the nanowire in the short-constriction limit~\cite{Kulik'70, Beenakker_PRL1991b} is mainly transmitted by the discrete electron-hole states (Andreev bound states) confined in the nanowire. These bound states appear because an electron incident on the semiconductor/superconductor interface is Andreev reflected as a hole with the opposite spin and momentum. To understand how Zeeman splitting and spin-orbit coupling affect Andreev bound states, it is instructive to consider two limiting cases: (a) magnetic-field-dominated regime where Zeeman splitting $V_z$ is much larger than the spin-orbit coupling $E_{\rm so}$. In this regime, propagating electron and hole modes are spin-polarized, and, thus, Andreev reflection is suppressed.
 (b) spin-orbit dominated regime $E_{\rm so} \gg V_z$. In this case, the propagating electron and hole have helical character and Andreev bound states can be formed by the helical electrons.
The intuition developed for the these two limits allows one to understand the general case of both strong Zeeman and SO couplings. We first consider short-junction limit and show that the helical phase in the semiconductor nanowires can be identified by a sharp decrease of the supercurrent as one increases an in-plane magnetic field. We also consider here long-junction limit and show that our findings persists in this case as well.

To have a better understanding of the Josephson transport in realistic situations, we consider the Andreev spectrum in disordered nanowires and calculate local density of states (LDoS) which can be probed in tunneling experiments. It is generally found that disorder smears out the peaks in the local density of states corresponding to the Andreev bound states and reduces transmission probability. However, we also show that magnetic field induced zero-energy crossings generically appearing in Andreev spectrum are robust against moderate disorder. This phenomenon appears when Andreev energy levels are non-degenerate due to the combination of spin-orbit and Zeeman couplings and is a consequence of the particle-hole symmetry which constraints certain matrix elements between a pair of zero-energy states at an isolated crossing. Thus, we predict that in this regime there will be generically zero-bias peaks in tunneling conductance as a function of the superconducting phase difference across the junction.

The paper is organized as follows: In Sec. I we present the theoretical model of the superconductor-nanowire-superconductor junction and the scattering matrix formalism for calculating the Josephson current. In Sec. II we examine in great detail the Andreev spectrum and Josephson current in short-junction limit. In Sec. III we study Josephson transport by numerically diagonalizing the Hamiltonian in a finite-size system which goes beyond the short-junction limit. In Sec. IV we consider the disorder effect.

\section{ Theoretical Model}\label{sec:model}
\begin{figure}[htb]
	\subfigure[]{
	\label{fig:junction:a}
	\begin{minipage}[t]{0.5\columnwidth}
		\includegraphics[width=\columnwidth]{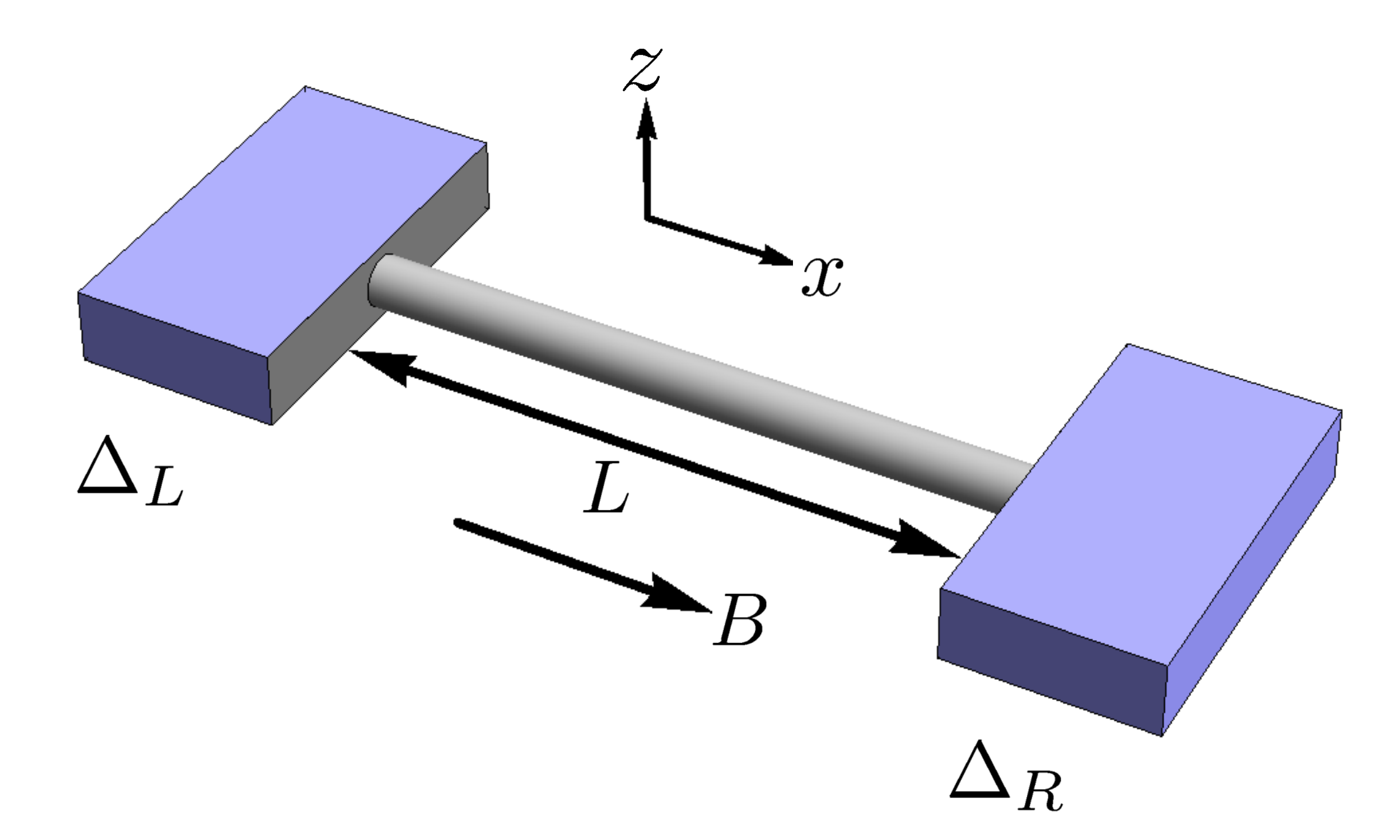}
	\end{minipage}}%
	\subfigure[]{
	\label{fig:junction:b}
	\begin{minipage}[t]{0.5\columnwidth}
 		\includegraphics[width=\columnwidth]{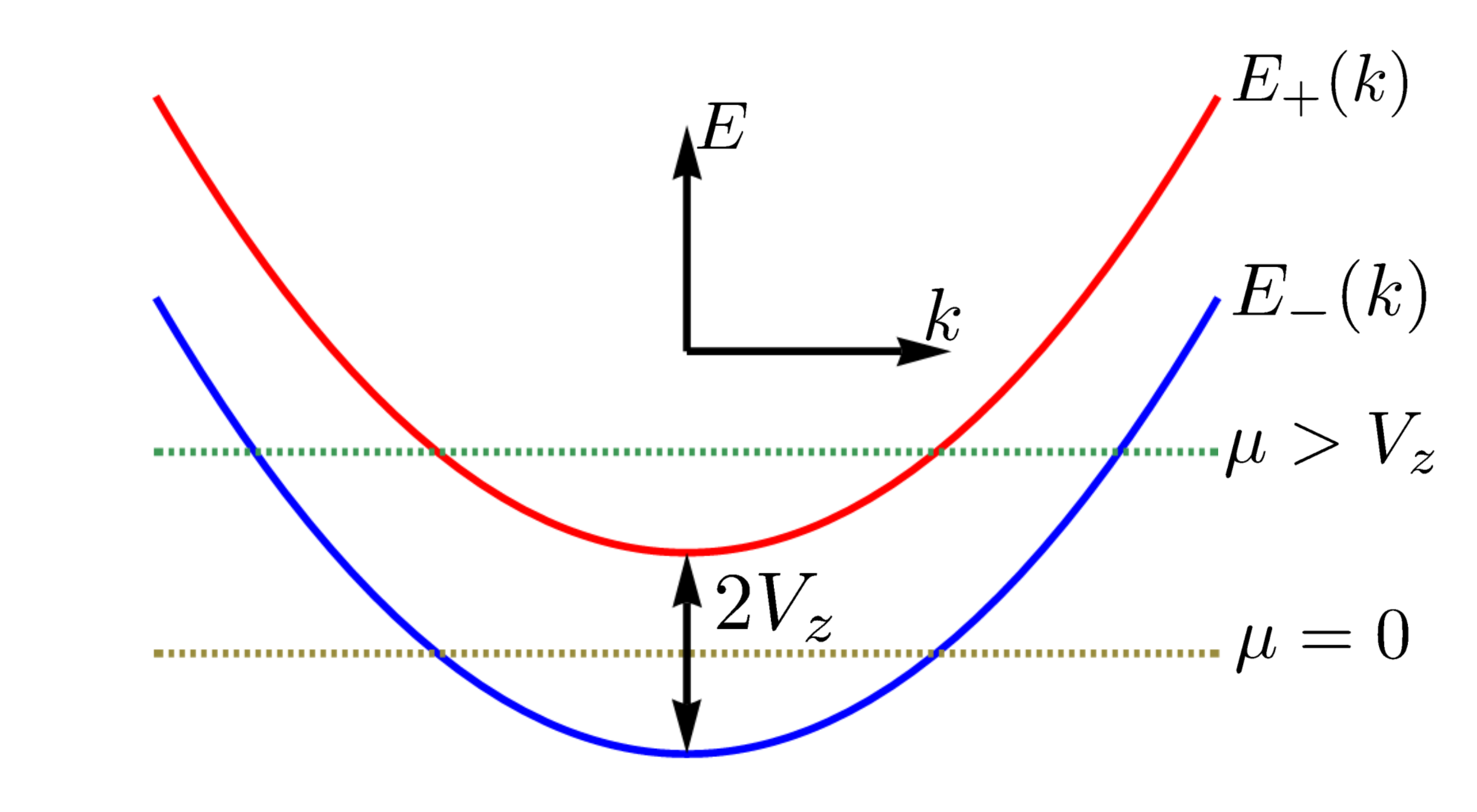}
	\end{minipage}
	}
	\caption{ (a) Schematic view of the SC-SM-SC junction. Semiconducting nanowire is coupled to s-wave superconducting electrodes. The chemical potential in the nanowire can be tuned by the gate (not shown). (b) Energy spectrum of the semiconductor with strong Rashba spin-orbit coupling and Zeeman splitting. The two values of the chemical potential at $\mu=0$ and $\mu > V_z$ correspond to helical and non-helical states of the nanowire, respectively.}
	\label{fig:junction}
\end{figure}

We consider SC-SM-SC junction shown in Fig.~\ref{fig:junction} where the single-channel semiconductor nanowire is modeled using the following Hamiltonian:
\begin{equation}
  \!\mathcal{H}_0\!=\!\psi^\dag(x)\!\Big( -\frac{\partial_x^2}{2m^*}\!-\!\mu\!+\!V_z\sigma_x\!-\!i\alpha\sigma_y\partial_x\Big)\!\psi(x),
	\label{eq:hamiltonian}
\end{equation}
where $m^*, \mu, \alpha$ are the effective mass, chemical potential and strength of Rashba spin-orbit coupling, respectively; $\psi$ is the spinor $(\psi_\uparrow,\psi_\downarrow)^T$ and we set $\hbar=1$. The magnetic field $B$ applied along the nanowire leads to the Zeeman energy $V_z=g_{\rm SM}\mu_B B/2$ where $g_{\rm SM}$ is the $g$-factor in the semiconductor which is assumed to be much larger than in the superconductor. One notices that the Hamiltonian is invariant under the combined operation $U=i\sigma_xP$ of spatial inversion $P: P f(x)\rightarrow f(-x)$ and spin $\pi$-rotation $i\sigma_x$. The eigenvalues of $H_{\rm SM}$ in momentum space are
\begin{equation}
	\varepsilon_\pm(k)=\frac{k^2}{2m^*}-\mu\pm\sqrt{\alpha^2 k^2+V_z^2}.
	\label{}
\end{equation}
Taking $\mu=0$ to be mid-way in the gap as a reference point, one can see that depending on whether $|\mu|$ exceeds $|V_z|$ or not, the semiconductor can have one or two Fermi surfaces. The phase with only one Fermi surface is called ``helical''  because the motion of an electron and its spin are correlated. Conversely, if $|V_z|<\mu$, there are two Fermi surfaces and we refer to this phase as non-helical (i.e. normal), see Fig. \ref{fig:junction:b}.

Two superconducting leads on both sides of the nanowire are described by s-wave BCS Hamiltonian with $\Delta_{R/L}=\Delta_0 e^{i \phi_{R/L}}$ being the corresponding order parameters, see Fig~\ref{fig:junction}(a). We assume that one can neglect Zeeman splitting in the superconductor due to the large difference in the Lande g-factors, {\it e.g.} the $g$-factor in InSb is $|g_{\rm SM}|\!\sim\!50$ whereas in the superconductor $g_{\rm SC} \approx 2$. Thus, one can tune the field to the regime where magnetic field induces large Zeeman splitting in the semiconductor without substantially suppressing superconductivity in the leads.

The quasiparticle spectrum in this system can be found by solving the Bogoliubov-de Gennes(BdG) equation:
\begin{equation}
  \begin{pmatrix}
	\mathcal{H}_0 & \Delta(x)\\
	\Delta^*(x) & -\sigma_y \mathcal{H}_0^*\sigma_y
  \end{pmatrix}\Psi(x)=E\Psi(x).
  \label{}
\end{equation}
Here $\Psi=(u_\uparrow, u_\downarrow, v_\downarrow, -v_\uparrow)^T$ is the wave function in the Nambu spinor space.

The dc Josephson current in the junction, as a function of the superconducting phase difference $\varphi$, is given by $I(\varphi)=-\frac{2e}{\hbar}\sum_{\varepsilon_n>0}\tanh\frac{\varepsilon_n(\varphi)}{2T}\odiff{\varepsilon_n(\varphi)}{\varphi}$ with
$\varepsilon_n$ being positive energy eigenvalues of Andreev states. Interestingly, the symmetry $U$ defined above implies that Andreev spectrum should be symmetric under $\varphi\rightarrow-\varphi$, provided the two interfaces are mirror symmetric, which is assumed in our calculations. 
If magnetic field and the direction of the spin-orbit coupling are not perpendicular to each other (i.e., there is a nonzero parallel component), the inversion symmetry with respect to left and right leads is completely broken and the spectrum can be asymmetric, leading to a spontaneous formation of supercurrent at the junction, see Ref.\onlinecite{Buzdin_PRL2008} for details.

We now consider the short-junction geometry corresponding to the length of the nanowire $L$ such that the dwell time of electrons $\tau_{\rm dw}$ in the nanowire $\tau_{\rm dw}\sim L/v_F \ll \Delta_0^{-1}$, which means that $L\ll \xi = \frac{v_F}{\Delta_0}$. In this case, the main contribution to the Josephson current comes from the Andreev bound states formed in the junction~\cite{Kulik'70, Beenakker_PRL1991b}. The spectrum of the Andreev bound states in the nanowire can be obtained using scattering matrix formalism~\cite{Beenakker_Review}. For technical reasons, we insert thin ballistic normal-state leads (N) to the left and the right of the semiconducting nanowire allowing for a well-defined scattering problem~\cite{Beenakker_Review}. The spatial separation of Andreev and normal scattering simplifies the problem substantially and allows one to relate the Josephson current to normal-state scattering matrix for N-SM-N system.

The spectrum of Andreev states is determined by the following condition~\cite{Beenakker_Review}:
\begin{equation}
	\det[1-r_\text{he} S_\text{e}(\varepsilon) r_\text{eh} S_\text{h}(\varepsilon)]=0,
	\label{eqn:andreevlevel}
\end{equation}
where
\begin{equation}
	\begin{gathered}
		r_\text{A}=
		\begin{pmatrix}
			e^{i\varphi/2} & 0\\
			0 & e^{-i\varphi/2}
		\end{pmatrix}\\
r_\text{he}=\gamma r_\text{A},\: r_\text{eh}=\gamma r_\text{A}^*,\gamma=e^{-i\arccos\frac{\varepsilon}{\Delta_0}},
	\end{gathered}
	\label{}
\end{equation}
and $S_\text{e(h)}$ is the normal-state scattering matrix for electrons (holes). The particle-hole symmetry relates $S_\text{h}$ to $S_\text{e}$ as
$	S_\text{h}(\varepsilon)= \Theta S_\text{e}(-\varepsilon)\Theta^{-1}$,
where $\Theta=i\sigma_y K$ is the particle-hole operator for spin-$1/2$ electrons and $K$ is the operator of complex conjugation~\cite{Dimitrova2006}.
We further neglect the energy dependence of scattering matrices $S_\text{e}(\varepsilon)\approx S_\text{e}(0)$, which is justified in the short-junction limit. We will examine the validity of the short-junction approximation later. We assume perfect transparency at the superconductor-nanowire interface unless stated otherwise.

\section{Short-Junction Results}\label{sec:short}
\subsection{Zeeman-field dominated regime}
We consider first the case of a large Zeeman splitting $V_z \gg E_{\rm so}\equiv m^*\alpha^2$, and show that the Josephson current exhibits two remarkable features depending on the value of the chemical potential: in the non-helical regime $V_z < \mu$, supercurrent exhibits oscillations with the applied magnetic field whereas in the helical phase $V_z > \mu$ the supercurrent is significantly suppressed.

The normal electron scattering matrix can be found exactly in the $\alpha=0$ limit.
\begin{equation}
	S_e=
	\begin{pmatrix}
		-iB_\uparrow/A_\uparrow^* & 0 & 1/A_\uparrow^* & 0\\
		0 & -iB_\downarrow/A_\downarrow^* & 0 & 1/A_\downarrow^*\\
		1/A_\uparrow^* & 0 & -iB_\uparrow/A_\uparrow^* & 0\\
		0 & 1/A_\downarrow^* & 0 & -iB_\downarrow/A_\downarrow^*
	\end{pmatrix},
	\label{}
\end{equation}
where the coefficients $A_\sigma,B_\sigma$ are defined as:
\begin{equation}
	\begin{split}
		A_\sigma&=\cos k_\sigma L+\frac{i}{2}\left( \frac{k_F}{k_\sigma}+\frac{k_\sigma}{k_F} \right)\sin k_\sigma L\\
		B_\sigma &=\frac{1}{2}\left( \frac{k_F}{k_\sigma}-\frac{k_\sigma}{k_F} \right) \sin k_\sigma L.
	\end{split}
	\label{eqn:AB}
\end{equation}
Here $k_F$ and $k_\sigma=\sqrt{2m(\mu-\sigma V_z)}$ are the Fermi momentum in the normal leads and electron momentum in the nanowire, respectively.

Having particle and hole scattering matrices, the spectrum of Andreev levels can be obtained using Eq.(5) of the main text. In this case of $V_z\neq 0, \alpha=0$, the spectrum is determined by the following equation:
\begin{equation}
	2\arccos\frac{\varepsilon}{\Delta_0} = \arccos\frac{B_\uparrow B_\downarrow+\cos\varphi}{|A_\uparrow A_\downarrow|}\pm \arg\frac{A_\uparrow}{A_\downarrow}+2n\pi.
	\label{eqn:andreevnoso}
\end{equation}
Here $n\in\mathbb{Z}$.

Let us now consider the limit  $V_z\ll \mu$. In this case, the spectrum of Andreev levels to the leading order in $V_z/\mu$ becomes $\varepsilon(\varphi)\approx \pm\Delta_0\cos\Big( \frac{\varphi}{2}\pm \varphi_0\Big)$ with $\varphi_0=V_zL/v_F$, and the Josephson current is given by
\begin{equation}
	\frac{I(\varphi)}{2e\Delta_0}=
	\begin{cases}
		\cos\varphi_0\sin\frac{\varphi}{2} & 0<\varphi<\pi-2\varphi_0\\
		-\sin\varphi_0\cos\frac{\varphi}{2} & \pi-2\varphi_0<\varphi<\pi+2\varphi_0\\
		-\cos\varphi_0\sin\frac{\varphi}{2} & \pi+2\varphi_0<\varphi<2\pi
	\end{cases},
	\label{eqn:Ic0a}
\end{equation}
 The current $I(\varphi)$ exhibits a jump at $\varphi=\pi\pm 2\varphi_0$, corresponding to the zero-energy crossings in the Andreev spectrum. The energy spectrum and the position of the crossings can be understood from the doubly-degenerate Andreev spectrum without any Zeeman field $\varepsilon(\varphi)=\pm\Delta_0\cos(\varphi/2)$, with zero-energy crossing at $\varphi=\pi$. The Zeeman field splits the degeneracy and the zero-energy crossings move away from $\pi$ by an amount $\varphi_0$. Note that zero-energy crossings are not robust and appear due to our assumption of perfect interface transparency ${\cal T}=1$.
 In realistic conditions ${\cal T}\neq 1$, and zero-energy crossings will become avoided crossings corresponding to the local minima of the energy spectrum and therefore vanishing of the Josephson current, which are accompanied with the sign reversal of the supercurrent~\cite{BobkovaPRB2002}, see Fig.\ref{fig:phase}.

The physical origin of the phase shift $\varphi_0$ can be understood quasi-classically from the Bohr-Sommerfeld quantization condition.  Indeed, Andreev bound states can only form when the total phase acquired by the electron undergoing two Andreev reflections is quantized:
$\oint p\di x+\Omega=2\pi m, m\in \mathbb{Z}.$
Here $\Omega=\pm\varphi-2\arccos\frac{\varepsilon}{\Delta}$ is the total scattering phase acquired by the electron (hole) at the interfaces and $\oint p\di x=\pm(k_\uparrow-k_\downarrow)L\approx \pm 2V_zL/v_F$. The latter agrees perfectly with the scattering matrix calculation. As follows from this analysis, the Josephson current in the limit of $V_z\ll \mu$ should oscillate with the Zeeman field $V_z$ with period which is proportional to $1/L$. Similar oscillations of the critical current have been found in SC-ferromagnet-SC junction~\cite{Buzdin_JETPL1982, pi-junction}.

In the polarized limit ($V_z \gg \mu$), the propagation of the spin-up electrons in the nanowire is exponentially suppressed. The calculation of the Josephson current in this limit yields the following result:
\begin{equation}
  I(\varphi)\approx {2e\Delta_0}e^{-\kappa L}\Big(1-\frac{\mu}{8V_z}\sin^2 k_\downarrow L\Big)\sin\varphi,
	\label{eqn:Ic0b}
\end{equation}
where $\kappa=\sqrt{2m^*(V_z-\mu)}$ and $k_\downarrow = \sqrt{2m^*(V_z+\mu)}$. Thus, if
the length of the wire $\xi \gg L \gg \kappa^{-1}$, the supercurrent through the nanowire will be significantly suppressed. This result is strictly speaking only valid when $\alpha=0$ as the presence of spin-orbit coupling leads to spin flips. However, the probability for the spin flips to occur within the distance $\kappa^{-1}$ is small because $E_{\rm so} \ll V_z$. Thus, the suppression of the supercurrent in the helical phase should be observable even in the presence of the small SOC as we explicitly demonstrate below.

\subsection{SOC dominated regime}
We now discuss Josephson effect in the case of small Zeeman splitting $V_z\ll E_{\rm so}$. The effect of magnetic field is perturbative here and we begin with $V_z=0$ limit.  Without any loss of generality, we can choose the direction of the spin-orbit coupling to be along $\hat z$-axis. The s-wave superconductor is spin-$\mathbb{SU}(2)$ invariant. Thus, the Hamiltonian for the nanowire reads
\begin{equation}
	H=\frac{p^2}{2m^*}-\mu-\alpha p\sigma_z.
	\label{}
\end{equation}
The energy spectrum for spin-up(-down) electrons is shifted to the right (left). Assuming that the chemical potential $\mu>0$, we can write down the wavefunction in the nanowire as
\begin{equation}
  \psi(x)=\sum_{\sigma=\uparrow,\downarrow}(b_\sigma^+ e^{ik_\sigma}+b_\sigma^- e^{-ik_\sigma})\ket{\sigma}.
\end{equation}
Here $k_{\sigma}$ are defined as
\begin{equation}
	k_{\sigma}=k_0+m^*\alpha \sigma, k_0=\sqrt{2m^*\mu+(m^*\alpha)^2}.
	\label{}
\end{equation}

Matching the boundary conditions at $x=\pm L/2$, we find the normal state scattering matrix to be
\begin{equation}
	S_e=\frac{1}{A^*}
	\begin{pmatrix}
		iB & 0 & e^{-im^*\alpha L} & 0\\
		0 & iB & 0 & e^{im^*\alpha L} \\
		e^{im^*\alpha L} & 0 & iB & 0\\
		0 & e^{-im^*\alpha L} & 0 & iB
	\end{pmatrix}.
	\label{}
\end{equation}
Here
\begin{equation}
  \begin{gathered}
	A=\cos k_0L +\frac{i}{2}\left(\frac{k_F}{k_0}+\frac{k_0}{k_F}\right)\sin k_0L\\
	B=\frac{1}{2}\left(\frac{k_0}{k_F}-\frac{k_F}{k_0}\right)\sin k_0L.
  \end{gathered}
	\label{}
\end{equation}
and $k_F$ is the Fermi momentum in the normal leads.

The spectrum of Andreev levels follow directly:
\begin{equation}
	 \varepsilon(\varphi)=\pm\Delta_0\sqrt{1-\frac{\sin^2(\varphi/2)}{1+\lambda^2\sin^2k_0L}}.
	\label{}
\end{equation}
Here $\lambda=\frac{1}{2}(\frac{k_F}{k_0}-\frac{k_0}{k_F})$ and $k_0=\sqrt{2m^*\mu+{m^*}^2\alpha^2}$, $k_F$ is the Fermi momentum in the superconductor. Andreev levels are doubly degenerate due to the presence of the time-reversal symmetry as well as the approximation $S_e(\varepsilon)\approx S_e(0)$ used here. The small corrections to the spectrum due to the presence of the spin-orbit coupling were considered in Refs.[\onlinecite{Nazarov'03,Beri'08}].
The main effect of spin-orbit coupling here is actually the modification of the Fermi momenta which affects normal reflection and thus the Josephson current due to the mismatch of the Fermi momenta in the nanowire and superconductor. For special values of $L$ satisfying the resonance condition $k_0L=n\pi, \, n\in\mathbb{Z}$, transmission of electrons through the spin-orbit coupled wire becomes perfect and the supercurrent is the same as in the absence of SOC.

\subsection{ General case}
We now turn to the general case when both SO and Zeeman terms are large. We can develop some intuition by considering Andreev reflection in terms of the eigenstates $\ket{\pm}$ of the Hamiltonian \eqref{eq:hamiltonian}. From the discussion of the Zeeman-dominated regime, we recall that there are two major consequences caused by Zeeman polarization: (a) the oscillation of the critical current with $V_z$ and $L$. (b) Significant suppression of the critical current when $V_z>\mu$.  However, the spin-orbit interaction causes precession of the spin of electrons/holes within the characteristic length scale $l_\text{so}$, which tends to wash out the spin polarization of electrons and holes. To this end, it is instructive to rewrite the s-wave pairing in the chiral basis:
\begin{equation}
	\Delta \psi_{\uparrow}^\dag(k)\psi_{\downarrow}^\dag(-k)=\Delta\sum_{\lambda,\lambda'=\pm}\chi_{\lambda\lambda'}(k)\psi_\lambda^\dag(k)\psi_{\lambda'}^\dag(-k),
	\label{}
\end{equation}
where $\chi_{\lambda\lambda}(k)=\alpha k/\sqrt{V_z^2+\alpha^2 k^2}$ and $\chi_{+-}(k)=V_z/\sqrt{V_z^2+\alpha^2 k^2}$.
Thus, both inter-band and intra-band pairing are generated in the chiral basis. Therefore, one expects that Andreev scattering at the N-S boundary has qualitatively similar behavior, i.e. there is both inter- and intra-band scattering.  We can understand our analytical results in the two limiting cases as follows: first in the Zeeman-dominated regime inter-band scattering dominates over intra-band scattering. Once the $\ket{+}$ band is gapped, quasiparticles in this band cannot propagate in the junction whose length is much larger than the length scale associated with the Zeeman gap. Therefore, the supercurrent will be suppressed, similar to the spin-polarized case discussed above.  In the opposite regime $E_\text{so}\gg V_z$, the Andreev reflection is dominated by the intra-band scattering and the situation is similar to the SOC-dominated regime where there is no abrupt change in the Josephson current across the boundary $\mu=V_z$.

The discussion above provides some intuition about Andreev reflection. However, in order to understand the coherent transport quantitatively, one needs to consider also normal reflection at the SC-SM interface. This can be done by appropriately matching the boundary conditions and finding the corresponding scattering matrix $S(\varepsilon)$ as described in the appendix. The expression of Josephson current in the general case is not particularly enlightening and we present here numerical results instead. We proceed by first re-scaling all the lengths and energies in units of $l_{\rm so}=\hbar^2/m^*\alpha$ and $E_{\rm so}=m^*\alpha^2/\hbar^2$, respectively. Using typical parameters for InAs $m^*\approx 0.04m_e, \alpha\approx 0.1\, \text{eV}\cdot\text{\AA}$, we estimate $l_\text{SO}\sim 100\, \text{nm}, E_\text{SO}\sim 1\text{K}$. The Josephson current is measured in units of $2e\Delta_0/\hbar$. Henceforth, all the re-scaled physical quantities are denoted by tilde.
\begin{figure}
	\begin{center}
		\includegraphics[width=\columnwidth]{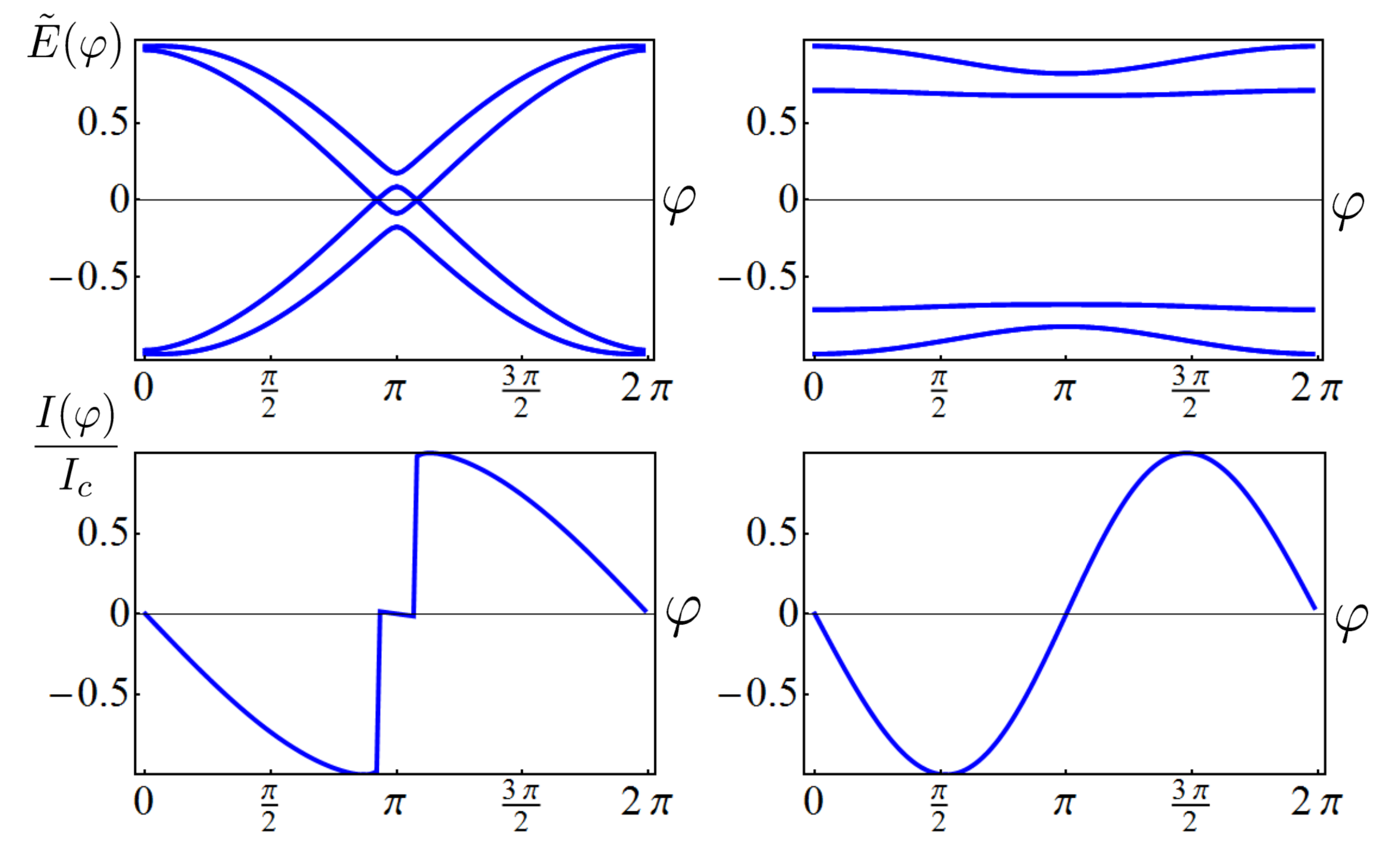}
	\end{center}
	\caption{ $I(\varphi)/I_c$ and $E(\varphi)$ in the non-helical phase (left panel) and the helical phase (right panel). The parameters are $\tilde{V}_z=0.1$, $\tilde{\mu}=0.5$ in the non-helical phase and $\tilde{\mu}=0.01$ in the helical phase.}
	\label{fig:phase}
\end{figure}

The Andreev spectra and current-phase relation for the SC-SM-SC junction are shown in Fig. \ref{fig:phase}.  In the non-helical phase, Zeeman field splits the degenerate Andreev levels and results in zero-energy crossings. As discussed above, finite transparency of the SM-SC interface will lead to avoided level crossings at these points. One can see that as we change chemical potential from non-helical to helical phase, the dispersion of Andreev levels with $\varphi$ decreases resulting in the suppression of the Josephson current.

\begin{figure}[b]
	\begin{center}
	\includegraphics[width=\columnwidth]{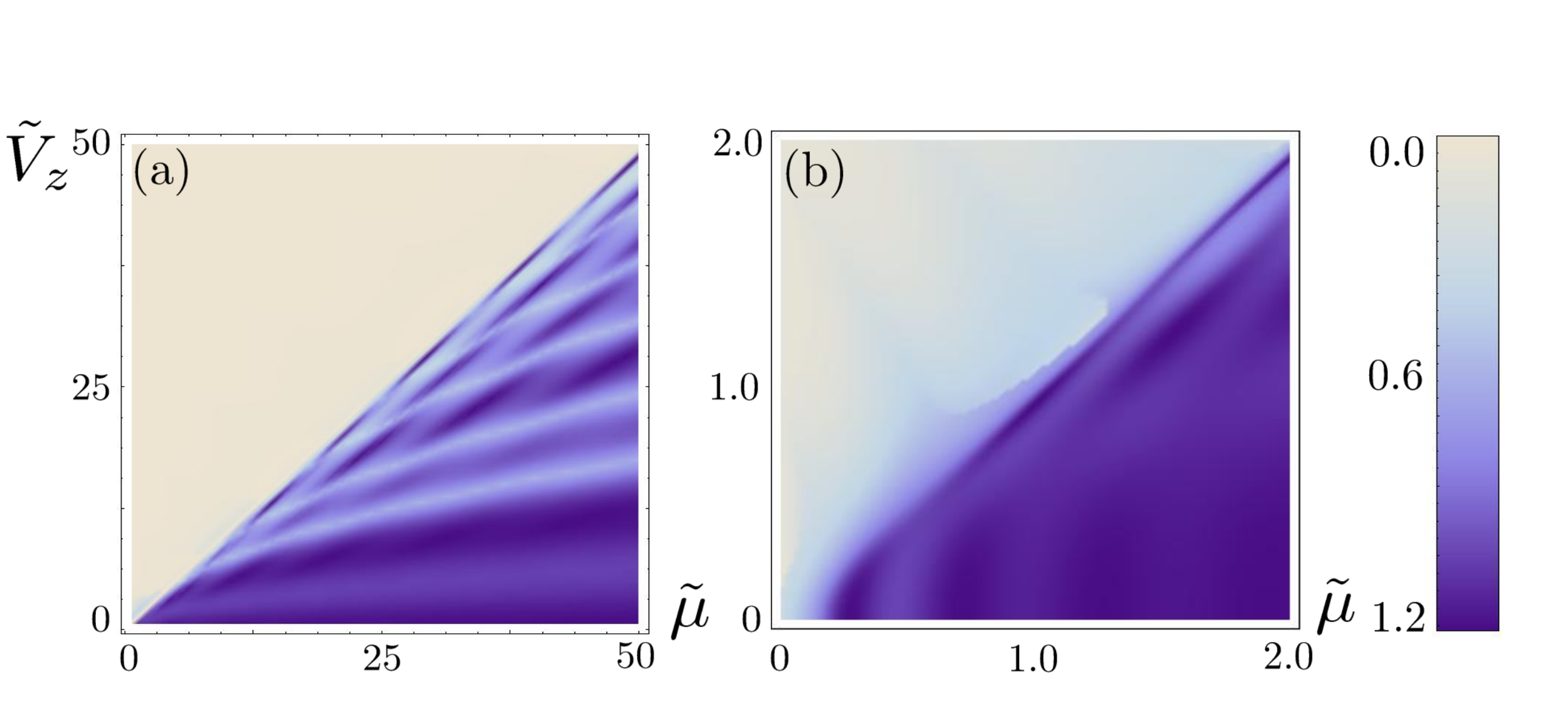}
\end{center}
\caption{Density plot of the current $|I(\pi/2)|$ as a function of $\tilde{V}_z$ and $\tilde{\mu}$. In order to understand the magnitude of the critical current, we need to decouple the phase dependence of the Josephson current which is non-trivial, especially around $\phi=\pi$. We find that it is more useful to plot $I(\varphi=\pi/2)$ instead of $I_c$. In Fig. (a) we show $|I(\pi/2)|$ for junction length $L=10$. Fig. (b) is a zoom-in of Fig. (a) for small $\tilde{V}_z$ and $\tilde{\mu}$ (or equivalently, large $E_{so}$). }
	\label{fig:current}
\end{figure}

The dependence of the Josephson current on chemical potential and Zeeman splitting is summarized in Fig. \ref{fig:current} in which we plot $|I(\pi/2)|$ as a measure of the size of the critical current. For $V_z \gg E_{so}$, the suppression of the Josephson current between helical and non-helical phases manifests as a sharp borderline $\tilde{V}_z=\tilde{\mu}$ in Fig.\ref{fig:current}(a), which actually persists to $V_z\sim E_{so}$. As one further increases SOC to $E_{so} \gg V_z$, such suppression is smeared out as shown in Fig.\ref{fig:current}(b). The critical current oscillations with magnetic field are also reflected in the ``stripes'' in the region $\tilde{V}_z<\tilde{\mu}$. One can notice that as $\mu$ increases the onset of the oscillations requires a larger $V_z$. Thus, the main features discussed qualitatively above within the simple models are also present in the general case.

\section{Long-junction results}\label{sec:long}
So far we have discussed Josephson effect for $L\ll \xi$. We now present results for the long-junction obtained using finite-size lattice calculation which involves full diagonalization of the BdG equations. This calculation is suitable for studying the Josephson effect beyond the short-junction limit allowing one to take into account the contribution to the current from the continuum spectrum. The tight-binding Hamiltonian reads
\begin{align}
    \!\!H&=\sum_{\langle ij\rangle,\sigma,\sigma'} [\delta_{\sigma,\sigma'}t_{ij} c_{j\sigma}^\dag c_{i\sigma'}+\alpha_ic^\dag_{j,\sigma}(\tau_x)_{\sigma\sigma'}c_{i\sigma'}+\text{h.c.}]\nonumber\\
\!\!&\!+\!\sum_i (V_{zi}\sigma\!-\!\mu_i)c^\dag_{i\sigma} c_{i\sigma}
  \!+\!\sum_i\!(\Delta_i c^\dag_{i\uparrow}c^\dag_{i\downarrow}+\text{h.c.}).
  \label{eqn:lattice}
\end{align}
Here we allow $V_z,\mu, \alpha, \Delta$ to be position-dependent to model the junction. $\bm{\tau}$ are the Pauli matrices in spin space. We take hopping amplitudes to be $t_{ij}=-t(\delta_{i,j-1}+\delta_{i,j+1})$. We assume that $\alpha_i=\alpha$ and $V_{zi}=V_z$ in the wire and zero otherwise; the SC pairing potential $\Delta_i$ is chosen to be non-zero outside of the wire with the magnitude  $\Delta_0=0.05t$ which corresponds to the coherent length $\xi\sim 10a$. In the numerical calculations, we take the length of the superconductors on both sides to be $L_\text{SC}=200 a$ and measure the energies and length in units of $t$ and $a$, respectively.

\begin{figure}
	\begin{center}
		\includegraphics[width=\columnwidth]{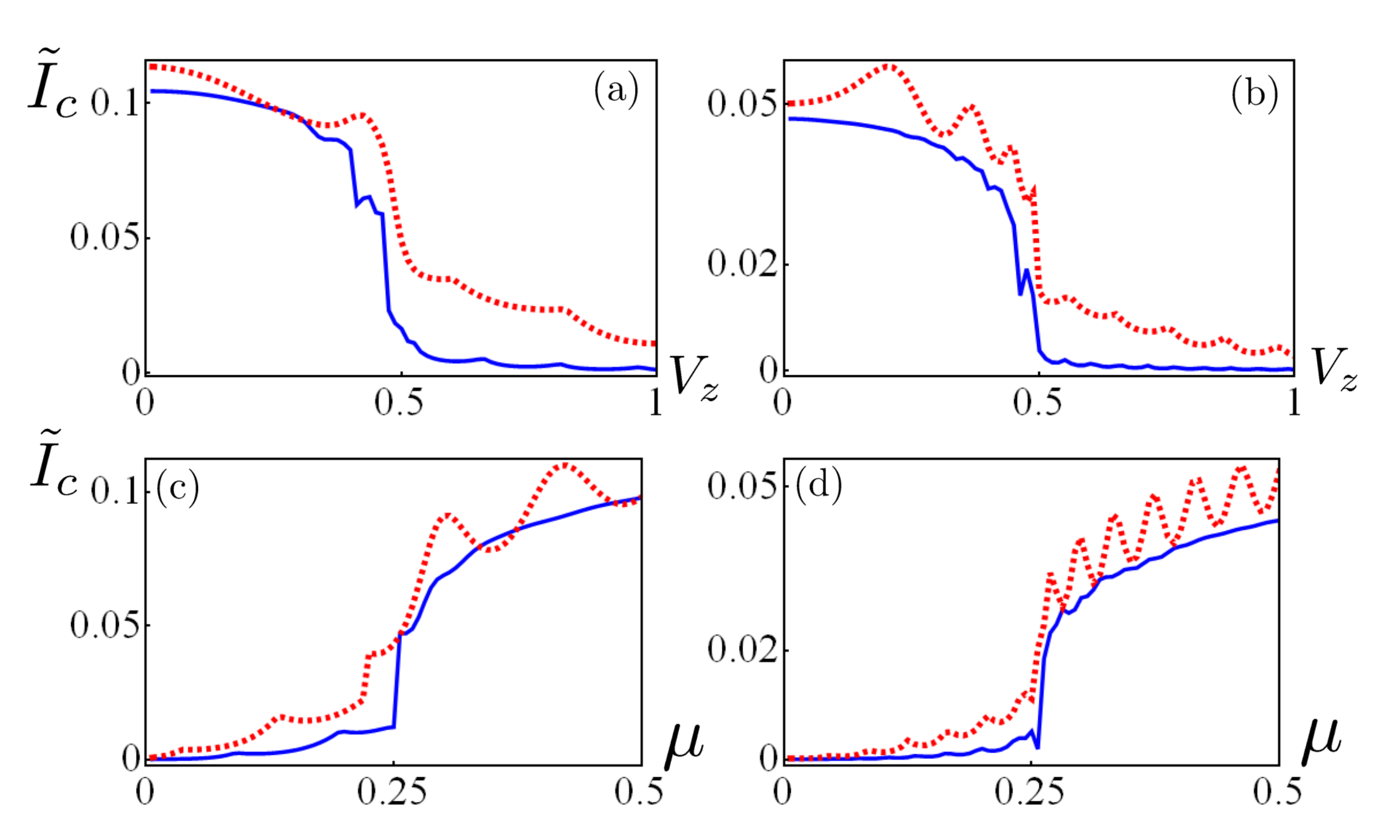}
	\end{center}
	\caption{Dependence of $\tilde{I}_c=|I(\pi/2)|$ on $\mu$ and $V_z$ from finite-size lattice diagonalization. The blue(solid) lines are results for $\alpha=0.5$ and the red(dashed) lines are results for $\alpha=1.5$. The parameters for each subfigure are: (a) $\mu=0.5, L=20$. (b) $\mu=0.5, L=60$. (c) $V_z=0.25, L=20$. (d) $V_z=0.25, L=60$. Here the the current is normalized by the corresponding value at $V_z=0$. Here $\mu$ is measured from the band bottom.}
	\label{fig:finite_mu}
\end{figure}

Having the excitation spectrum of the system as a function of the phase difference $\varphi$, one can compute Josephson current through the nanowire. The dependence of the Josephson current on $\mu$ and $V_z$ is shown in Fig. \ref{fig:finite_mu}. The suppression of the Josephson current as one sweeps parameters between helical and non-helical phases ($\mu\approx V_z$) is also visible in the long-junction limit for small values of the SOC, see results for $\alpha=0.5$ in Fig. \ref{fig:finite_mu}. As SOC is increased to $\alpha=1.5$, the sharp suppression of the current is smoothen out and the difference in the critical current between the helical and non-helical phase becomes less significant. Overall, we find that the interesting features in the Josephson effect discussed above persist in the long-junction limit.

\section{Local Density of States at the junction}\label{sec:LDOS}

In addition to measuring the Josephson current discussed in the previous sections, one can also extract information about Andreev energy spectrum using tunneling spectroscopy experiments, effectively measuring local Density of States (LDoS). In these experiments, it is important to take into account effect of disorder since it will lead to a reduction of the transmission probability and, thus, will strongly affect the energy spectrum. For example, in the SNS system in the short junction limit, the zero-energy crossing at $\varphi=\pi$ as evident from Andreev spectrum $\varepsilon(\varphi)=\pm\Delta\cos\frac{\varphi}{2}$ is removed once we introduce impurities which lead to the reduction of the junction transparency\cite{Beenakker_Review}. The situation is different in semiconductor nanowires in a magnetic field as we explain below.

\begin{figure}[htb]
  \begin{center}
	\includegraphics[width=\columnwidth]{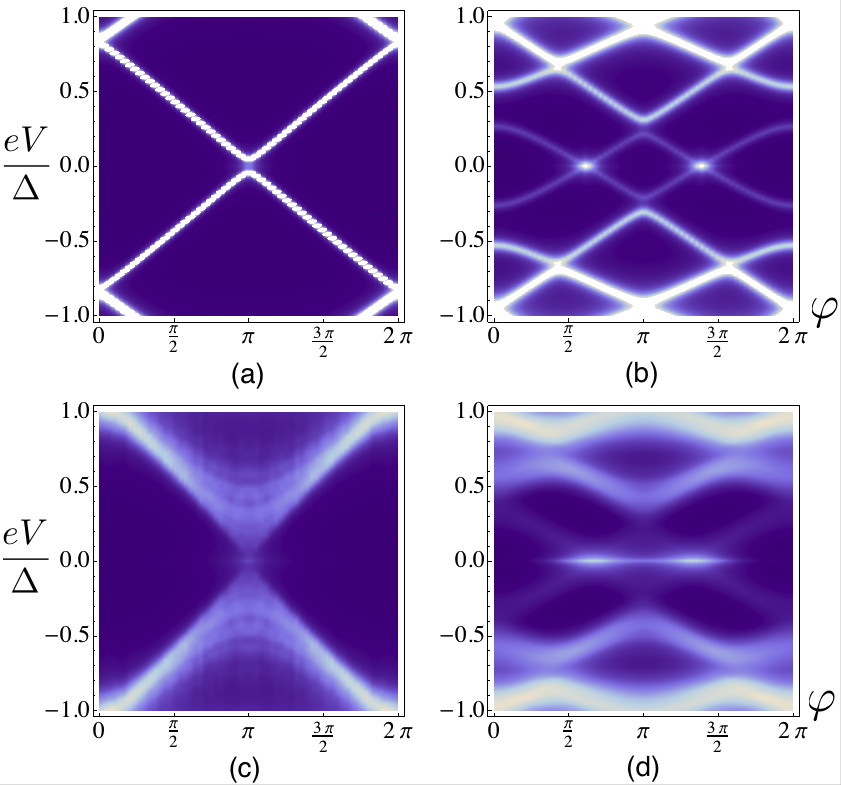}
  \end{center}
  \caption{LDOS measured in the middle of the junction. In both plots, $\Delta_0=0.05, L=20, \alpha=0.5, \mu=0.5$. (a) $V_z=0, W=0$ (b) $V_z=0.4, W=0$ (c) $V_z=0, W=0.3$. (d) $V_z=0.4, W=0.3$. }
  \label{fig:ldos}
\end{figure}

Let's consider LDoS defined as
\begin{equation}
  \rho(\vr,\omega)=\sum_{n}\big[|u_{n\sigma}(\vr)|^2\delta(\omega-E_n)+|v_{n\sigma}(\vr)|^2\delta(\omega+E_n)\big],
  \label{}
\end{equation}
Here $u_\sigma, v_\sigma$ are eigenstates of the BdG equation corresponding to eigenenergy $E_n$. In our numerical calculations, we use Lorentzian function instead of the $\delta$ function and introduce some broadening of the energy levels.
The LDoS in the clean system is shown in Fig.\ref{fig:ldos}a) and b). As discussed above, magnetic field generically leads to the appearance of the zero-energy crossings in Andreev spectrum. However, unlike in SNS junction, these crossings are robust against disorder which is a distinct signature of semiconductor nanowires in a magnetic field where Andreev states are non-degenerate in contrast with the conventional normal metals having spin-degenerate spectrum. These differences can be seen in LDoS measurements in a magnetic field which can be done experimentally since LDoS is directly related to the local differential tunneling conductance:
\begin{equation}
  \frac{\di I(\vr,eV)}{\di V}\propto \rho(\vr, \omega=eV).
  \label{}
\end{equation}

We now discuss the disorder-averaged LDoS $\rho(\vr,\omega)$ where these features are visible. The disorder can be included in the lattice calculations by adding an on-site random chemical potential to the Hamiltonian~\eqref{eqn:lattice}:
\begin{equation}
  H_{\text{disorder}}=\sum_i V_i c_i^\dag c_i.
  \label{}
\end{equation}
The random potential $V_i=V(\vr_i)$ satisfies Gaussian distribution with the correlation function $\langle V(\vr)V(\vr')\rangle_{\text{dis}}=W^2\delta(\vr-\vr')$. In conventional metals, the strength of the disorder potential determines the electron mean-free path $l=v_F\tau$ with $\tau^{-1}= 2\pi W^2\nu(\varepsilon_F)$ which is the same for both spins. Here $\nu(\varepsilon_F)$ is the normal-state density of states at Fermi level, $\nu(\varepsilon_F)=\frac{1}{\pi}(\frac{\di \varepsilon(k)}{\di k})^{-1}|_{k=k_F}$. In the presence of spin-orbit coupling and Zeeman field, electrons in different bands have different mean-free paths due to, for example, Fermi velocity mismatch. In addition, the singularity of the DoS at the bottom of the band leads to strong energy dependence of the mean-free path. Thus, unlike in conventional metals $l$ is not a good quantity to characterize the strength of the disorder.  However, for the lack of a better quantity to characterize the disorder, we roughly estimate $l\approx 5$ for the parameters used in the numerical calculation and away from the bottom of the band. Thus, in our simulations $l$ is much smaller than the length of the junction $L=20$ and, thus, effect of the disorder is quite important here.

For the sake of concreteness,  we now consider LDoS in the short-junction limit in the presence of spin-orbit coupling. From the scattering matrix calculation in Sec.\ref{sec:short}, we already know that the spectrum of Andreev bound states has no zero-energy crossings, see Fig.~\ref{fig:ldos}(a). It is naturally expected that disorder does lead to any qualitative changes in this case apart from broadening the peaks in the LDoS and increasing the gap at $\varphi=\pi$. This is confirmed by the numerical simulations shown in Fig.~\ref{fig:ldos}(a) and (c). The Zeeman field splits the spin $\mathbb{SU}(2)$ degeneracy and results in a complicated spectrum as demonstrated in Fig.\ref{fig:ldos}(b) for the clean case and Fig. \ref{fig:ldos}(d) for the disordered case. One can notice that there are two magnetic field-induced zero-energy crossings in the spectrum of Andreev bound states on the interval $\phi=[0,2\pi]$. From the disorder-averaged LDoS shown in Fig. \ref{fig:ldos}(d), we see that the crossing at zero energy persists at least for moderate disorder strength. The reason of this robustness is due to particle-hole symmetry as we explain below.

 Let's consider an isolated crossing in the energy spectrum for certain superconducting phase difference $\varphi=\varphi_0$. In the spirit of $k\cdot p$ perturbation theory, one can analyze effect of various perturbations around this level crossing. The two zero-energy states $\Psi_0$ and $\Psi_1$ at the crossing are related by particle-hole symmetry, namely $\Psi_1=\Xi\Psi_0^*$ where $\Xi=\sigma_y\tau_y$ in the Nambu spinor notation adopted here. To understand the stability of the crossing, we consider now an arbitrary local perturbation $O(x)$ (e.g. disorder potential) and calculate its matrix element between zero-energy states. The perturbation $O(x)$ has to satisfy particle-hole symmetry $\Xi^\dag O(x)\Xi=-O^T(x)$.
Using this relation, one an show that the off-diagonal matrix element $O_{10}=\tpd{\Psi_1}{O}{\Psi_0}$ is zero. Indeed, one can show that
\begin{equation}
  \begin{split}
  O_{10}&=\tpd{\Psi_1}{O}{\Psi_0}=\tpd{\Psi_0^*}{\Xi^{\dag}O}{\Psi_0}\\
  &=-\tpd{\Psi_0^*}{O^T\Xi^\dag}{\Psi_0}\\
  &=-\int\di x\,\Psi_0^T(x) O^T(x)\Xi^\dag\Psi_0(x)
\end{split}
  \label{}
\end{equation}
Since $\Xi^T=\Xi$, one finds that $\delta O_{10}^T=-\delta O_{10}=0$. Thus, any on-site perturbation preserving particle-hole symmetry does not mix $\Psi_0$ and $\Psi_1$ states. Another way of proving this fact is to show that the two zero-energy levels actually correspond to different fermion parity\cite{Fu_PRB2009} and as such are robust as long as there are no other zero-energy crossings nearby. However, generic perturbation produces non-zero diagonal contributions $\tpd{\Psi_0}{O}{\Psi_0}=-\tpd{\Psi_1}{O}{\Psi_1}$ which shift the position of the crossing. Thus, in the subspace spanned by the zero-energy eigenstates, an on-site perturbation acts as $\mu_z$.

We now apply these results to the zero-energy crossings in the Andreev bound spectrum. As discussed in Sec:\ref{sec:short}, for $V_z<\mu$ there might be two such crossings in the spectrum located symmetrically with respect to $\varphi=\pi$. Each crossing has a pair of zero-energy states since spin degeneracy is removed by the magnetic field. One of the relevant perturbation near the crossing point is the deviation of the superconducting phase from $\delta \varphi$. According to our general argument, we can expand in small deviation of the phase difference $\Delta\varphi\equiv\varphi-\varphi_{0i}$ to obtain:
\begin{equation}
  \delta H_\varphi=(a\Delta\varphi+b\Delta\varphi^2+\cdots)\mu_z.
  \label{}
\end{equation}
Here $\mu_z$ is the Pauli matrix acting in the space spanned by the two zero-energy eigenstates. Other perturbations, such as on-site disorder, also take the form
\begin{equation}
  \delta H'=\delta O_{00}\mu_z.
  \label{}
\end{equation}
Combining the two contributions, we find that the spectrum now reads
\begin{equation}
  E=\pm\left|a\Delta\varphi+\delta O_{00}\right|.
  \label{}
\end{equation}
Therefore, the perturbation shifts the crossing point $\varphi_0\rightarrow \varphi_0+\frac{\delta O_{00}}{a}$. If the perturbation is strong enough, then the two crossings located at different $\varphi_i$ are brought close to each other and can annihilate each other since these perturbations can now mix the states corresponding to different crossings and, thus, can generically open a gap in the spectrum. However, such a perturbation needs to be larger than certain threshold which explains why zero-energy crossings survive in the presence of a moderate disorder.

\section{Conclusions and Discussions}
We have developed a theory for the coherent transport through semiconductor nanowire with strong spin-orbit coupling and Zeeman splitting. We find a number of remarkable features in the Josephson current through the nanowire originating from the interplay between two competing terms: spin-orbit coupling and Zeeman splitting. In particular, we show that the transition between helical and non-helical states of the nanowire is accompanied by a large suppression of the critical current through the junction when Zeeman energy is large compared to the spin-orbit energy scale. We support these findings by calculating the Josephson current both in the short-junction limit using scattering matrix formalism and in the long-junction case using the numerical diagonalization. The suppression of the critical current can be used as a diagnostic of the helical state in the nanowire. We also show that the critical current exhibits oscillations with an applied magnetic field. These oscillations are generic and might be relevant for recent experiments on topological insulators/supercoductor structures~\cite{Samarth'11}.

We also compute local density of states and study effect of disorder on the Andreev energy spectrum. Without an applied magnetic field, the disorder-averaged local density of states is qualitatively similar to that of the clean case - the disorder mainly changes the junction transparency. In the presence of a magnetic field removing spin-degeneracy of the Andreev energy spectrum, we find zero-energy crossings which are robust against moderate disorder. We explain this robustness by invoking general arguments based on particle-hole symmetry and show analytically that any on-site perturbation only shifts the position of an isolated zero-energy crossing but does not open a gap in the spectrum.

\section{ Acknowledgements}
We thank Liang Fu, Leo Kouwenhoven, Yuval Oreg, Yaroslav Tserkovnyak, Nitin Samarth and Xiong-Jun Liu for stimulating discussions, and Shuo Yang for the help with the figures. RL acknowledges hospitality of the Aspen Center for Physics supported by NSF grant \#1066293. MC was supported by Microsoft Q, DARPA-QuEST and KITP Graduate Fellowship under NSF PHY05-51164 grant.
\appendix
\section{Derivation of the scattering matrix $S(\varepsilon)$}
\label{sec:appb}
We now briefly outline the derivation of the particle component of the scattering matrix $S_\text{e}(\varepsilon)$. The system we consider here consists of the semiconductor nanowire $-L/2<x<L/2$ coupled to the superconducting electrodes at $x=\pm L/2$. The particles propagate freely within the nanowire $-L/2<x<L/2$ in the presence of a finite spin-orbit coupling and Zeeman field (i.e. $\alpha\neq 0, V_z\neq 0$). For technical reasons, we insert thin ballistic normal leads to the left and right of the semiconducting nanowire~\cite{Beenakker_Review}. We then calculate particle (electron) scattering matrix by considering N-SM-N system where the scattering at the SM-N interface occurs, for example, due to the mismatch of the Fermi momenta. The scattering at the N-SC interface is assumed to be ideal(i.e. normal reflection-free) consisting of Andreev reflection at the SC-N interface. This spatial separation allows one to substantially simplify the problem and relate supercurrent to the normal-state scattering matrix.

Let us first construct a basis for the particle component of the total scattering matrix. The wave incident on the nanowire corresponds to a vector:
\begin{align}
	c^\text{in}=(c^+_{\uparrow}(L), c^+_\downarrow(L),c^-_{\uparrow}(R), c^-_\downarrow(R)).
\end{align}
The corresponding outgoing wave is given by
\begin{align}
	c^\text{out}=(c^-_{\uparrow}(L), c^-_\downarrow(L),c^+_{\uparrow}(R), c^+_\downarrow(R))
\end{align}
where $c^{\pm}_{\sigma}(L)$ correspond to right/left-moving quasiparticles with spin $\sigma$ in the normal region at the left boundary. The normal scattering matrix is defined as $c^{\text{out}}=S_e(\varepsilon) c^\text{in}$. The wavefunction in the N-SM-N system is given by
\begin{widetext}
\begin{equation}
	\psi(x)=
	\begin{cases}
		\displaystyle \sum_{\sigma=\uparrow,\downarrow}c^+_\sigma(L)\varphi_\sigma e^{ik(x+ L/2)}+c^-_\sigma(L)\varphi_\sigma e^{-ik(x+ L/2)} & x<-L/2\\
		\displaystyle\sum_{s=\pm}b^+_s \chi_s(k_s) e^{ik_s x}+b^-_s \chi_s(-k_s) e^{-ik_s x} & -L/2<x<L/2\\
		\displaystyle\sum_{\sigma=\uparrow,\downarrow}c^+_\sigma(R)\varphi_\sigma e^{ik (x-L/2)}+c^-_\sigma(R)\varphi_\sigma e^{-ik(x- L/2)} & x>L/2
	\end{cases}
	\label{}
\end{equation}
\end{widetext}
Here $k=\sqrt{2m(\varepsilon+\mu)}$ where $\mu$ is the chemical potential in the normal metal. The spinor wavefunction $\varphi_\sigma$ and $\chi_s(k)$ are given by
\begin{equation}
  \begin{gathered}
	\varphi_\uparrow =
	\begin{pmatrix}
		1\\ 0
	\end{pmatrix}, \varphi_\downarrow =
	\begin{pmatrix}
		0\\ 1
	\end{pmatrix},\\
	\chi_+(k)=
	\begin{pmatrix}
		\cos \theta_k\\
		\sin\theta_k
	\end{pmatrix},\,\chi_-(k)=
	\begin{pmatrix}
		-\sin\theta_k\\
		\cos\theta_k
	\end{pmatrix}.
  \end{gathered}
	\label{}
\end{equation}
Here $\tan 2\theta_k = \alpha k/V_z$.

At the right/left boundaries $x=\pm L/2$, the wavefunction and the probability current have to be continuous. For example, at $x=L/2$ we have the following boundary conditions:
\begin{equation}
	\begin{gathered}
	\psi(L/2+0^+)=\psi(L/2-0^+),\\
	\partial_x\psi(x)|_{x=L/2+0^+}=(\partial_x+im\alpha\sigma_x)\psi(x)|_{x=L/2-0^+}
\end{gathered}
	\label{}
\end{equation}
Matching the boundary conditions yields a system of equations for $c_\sigma^\pm, b_s^\pm$. For example, at $x=-L/2$ we find
\begin{widetext}
\begin{equation}
	\begin{gathered}
		c_\uparrow^+(L) + c_\uparrow^-(L)=(b_+^+e^{-ik_+L/2} + b_+^-e^{ik_+L/2})\cos{\theta_+}-(b_-^+e^{-ik_-L/2} - b_-^-e^{ik_-L/2})\sin{\theta_-}\\
		c_\downarrow^+(L) + c_\downarrow^-(L)=(b_+^+e^{-ik_+L/2} - b_+^-e^{ik_+L/2})\sin{\theta_+}+(b_-^+e^{-ik_-L/2} + b_-^-e^{ik_-L/2})\cos{\theta_-}\\
		k(c_\uparrow^+(L) - c_\uparrow^-(L))=(k_+\cos{\theta_+}+k_R\sin{\theta_+})(b_+^+e^{-ik_+L/2} - b_+^-e^{ik_+L/2})+(k_R\cos{\theta_-}-k_-\sin{\theta_-})(b_-^+e^{-ik_-L/2} + b_-^-e^{ik_-L/2})\\
		k(c_\downarrow^+(L) - c_\downarrow^-(L))=(k_R\cos{\theta_+}+k_+\sin{\theta_+})(b_+^+e^{-ik_+L/2} + b_+^-e^{ik_+L/2})+(k_-\cos{\theta_-}-k_R\sin{\theta_-})(b_-^+e^{-ik_-L/2} - b_-^-e^{ik_-L/2})
	\end{gathered}
	\label{}
\end{equation}
\end{widetext}
Here $k_R=m \alpha$. The equations at $x=L/2$ are similar. Eliminating $b_s^\pm$, we obtain the particle scattering matrix $S_e (\varepsilon)$. Using the particle-hole symmetry of the BdG equations one can obtain $S_h(\varepsilon)$ and compute the spectrum of Andreev levels using Eq.(5) of the main text.

\section{Validity of the short-junction approximation}

In the calculation of the Josephson current, we have made two approximations: (1) We assumed that the Josephson current is dominated by the Andreev bound states in the junction (i.e. short-junction limit), and neglected the contribution of the continuum states. (2) In evaluating the Andreev spectrum we have neglected the energy dependence of the normal state scattering matrix: $S_e(\varepsilon)\approx S_e(0)$. In this section we examine the validity of the second approximation. We start with the case $\alpha=0$ and show that such approximation is justified apart from the points close to $\varphi=0,2\pi$.

We first derive the equation determining the Andreev levels with full energy dependence of the scattering matrix $S_e(\varepsilon)$. When $\alpha=0$, the dispersion of the Andreev levels is given by
\begin{equation}
	\begin{gathered}
	2\arccos\frac{\varepsilon}{\Delta_0} = \pm\Big[\arccos\frac{B_\uparrow(\varepsilon) B_\downarrow(-\varepsilon)\!+\!\cos\varphi}{|A_\uparrow(\varepsilon) A_\downarrow(-\varepsilon)|} +\arg\frac{A_\uparrow(\varepsilon)}{A_\downarrow(-\varepsilon)}\Big],\\
		2\arccos\frac{\varepsilon}{\Delta_0} = \pm\Big[ \arccos\frac{B_\uparrow(-\varepsilon) B_\downarrow(\varepsilon)\!+\!\cos\varphi}{|A_\uparrow(-\varepsilon) A_\downarrow(\varepsilon)|} -\arg\frac{A_\uparrow(-\varepsilon)}{A_\downarrow(\varepsilon)}\Big],
\end{gathered}
		\label{eqn:level2}
\end{equation}
where $A_\sigma$ and $B_\sigma$ are defined in Eq.~\eqref{eqn:AB} with $k_\sigma\equiv k_\sigma(\varepsilon)=\sqrt{2m(\mu+\varepsilon-\sigma V_z)}$.

Assuming that both $V_z$ and $\Delta$ are much smaller than $\mu$, one can treat $V_z$ and $\varepsilon$ as small parameters and perform perturbative expansion of Eq.~\eqref{eqn:level2}. To the leading order in $\varepsilon/\mu$ and $V_z/\mu$, the coefficients $B_\sigma$ and $A_\sigma$ are given by:
\begin{equation}
	A_\sigma\approx e^{ik_\sigma(\varepsilon) L}, B_\sigma\approx-\frac{\varepsilon-\sigma V_z}{\mu}\sin k_\sigma L
	\label{}
\end{equation}
Substituting into \eqref{eqn:level2}, we obtain
\begin{widetext}
\begin{equation}
	\begin{split}
		2\arccos\frac{\varepsilon}{\Delta_0}&=\pm\Big\{\arccos \Big[\!-\!\frac{(\varepsilon\!-\!V_z)^2}{\mu^2}\sin^2 k_F L+\cos\varphi\Big]
		-\frac{2(\varepsilon-V_z)L}{v_F}\Big\}	\\
		2\arccos\frac{\varepsilon}{\Delta_0}&=\pm\Big\{\arccos \Big[\!-\!\frac{(\varepsilon\!+\!V_z)^2}{\mu^2}\sin^2 k_F L+\cos\varphi\Big]
		+\frac{2(\varepsilon+V_z)L}{v_F}\Big\}.
	\end{split}
	\label{eqn:andreev5}
\end{equation}
\end{widetext}
Compared to the short-junction limit, an extra term $\varepsilon L/v_F$ appears in the equation. As a rough estimate, one can put in $\varepsilon\sim \Delta_0$. Then, this term becomes $\Delta_0 L/v_F=L/\xi$. When $L/\xi\ll 1$, we can disregard this term and the short-junction approximation is justified.

However, in some cases that even for small $L/\xi$ this correction can result in a qualitatively different dependence on $V_z$. Let us consider the spectrum of Andreev levels at $\varphi\approx 0$. Since the time-reversal symmetry of the system is broken by the Zeeman term, we expect the Andreev levels to be split. However, if we start with the approximation of $S_e(\varepsilon)\approx S_e(0)$, we find
\begin{equation}
	\varepsilon(\varphi=0)=\pm \Delta_0\cos\Big(\frac{V_z\sin k_F L}{\mu}\pm\frac{V_zL}{v_F}\Big),
	\label{}
\end{equation}
which gives in the limit of $V_z \rightarrow 0$ the energy splitting quadratic in $V_z$:
\begin{equation}
	\delta\varepsilon \approx 2 \Delta_0 \left(\frac{V_z}{\mu}\right)^2 k_F L \sin k_F L.
	\label{}
\end{equation}
However, this scaling is qualitatively different if we include the energy dependence. Indeed, assuming $k_FL\gg 1$, we find that
\begin{equation}
	\delta\varepsilon \approx \left(\frac{L}{\xi}\right)^2 V_z
	\label{}
\end{equation}
as one would expect from qualitative considerations. Thus, apart from the points $\varphi\approx 0, 2\pi$, the approximation $S_e(\varepsilon)\approx S_e(0)$ gives qualitatively correct results.

Let us now discuss the case with a finite SOC. Generally, one has to rely on numerical solution but under the assumption that both $\alpha$ and $V_z$ are small with respect to $\mu$, we obtain the following analytical equation for the Andreev levels, which agrees with the one derived in Ref.~[\onlinecite{Bezuglyi_PRB2002}]:
\begin{equation}
	\arccos\frac{\varepsilon}{\Delta_0}=\frac{\varepsilon L}{v_F} +\frac{\varphi}{2}\pm\arcsin\Big[\frac{V_z|\sin( E L/v_F)|}{E}\Big]+n\pi.
	\label{}
\end{equation}
Here $E=\sqrt{V_z^2+\alpha^2k_F^2}$.
 However there is a subtle point here: at $\alpha=0$ the above equation reduces to \eqref{eqn:andreev5} for $\varphi\neq 0$ up to linear order in $V_z$. However, as explained above Eq.~\eqref{eqn:andreev5} around $\varphi=0$ contains an additional term $\frac{(\varepsilon\pm V_z)\sin k_FL}{\mu}$ which was neglected in Ref.~[\onlinecite{Bezuglyi_PRB2002}].


\end{document}